\documentclass[11pt]{article}
\usepackage{graphicx}
\def\title{\begin{center}\Large\bf}
\def\author(s){\vspace{0.3cm}\large\rm}
\def\text{\end{center}}

\begin{document}


\noindent { \small{\it
\begin{center}
    Contribution to the SPIE meeting in
Marseille on 27 June 2008\\
"Digital revival of intensity interferometry with atmospheric
Cherenkov Telescope arrays"
\end{center}
}}
\medskip \hrule \medskip

\bigskip
\bigskip


\title
Bispectral technique for reconstruction the astronomical images with
an intensity interferometer
\bigskip



\author(s)
B.E. Zhilyaev

\bigskip

\smallskip

\noindent {\small {\it Main Astronomical Observatory, NAS
of Ukraine,  27 Zabolotnoho, 03680 Kiev, Ukraine}} \\

\noindent {\small {\it e-mail:}} {\small{\bf zhilyaev@mao.kiev.ua}}

\smallskip


\text


\section*{\small {Abstract}}

\small{\bf An extension may be proposed to the intensity
interferometer of Hanbury Brown and Twiss to provide the Fourier
phase measurement by the use of third-order intensity correlations.
It is well known that interferometric reconstruction of astronomical
images can be obtained from second-order correlations only when {\it
a priory} information is used about the object. The third-order
intensity correlations contain information about the Fourier phase
and need no such assumptions. In the ordinary way we can make
measurements of the second-order intensity correlations with an
intensity interferometer. We can also calculate the third-order
intensity correlations from the data set of measured intensities for
each distance triplet. After that we can reconstruct the Fourier
phase from third-order correlations by using bipectral technique.
When this is combining with the Fourier magnitude obtained from the
second-order intensity correlations, we have the ultimate Fourier
transform of the source brightness distribution. An inverse Fourier
transform recovers the source image.}

\bigskip

According to the van Cittert-Zernike theorem any quasi-monochromatic
source creates in a projective plane spatial coherent pattern. It is
described by complex degree of coherence $\gamma(P_{1},P_{2})$ and
is equal to the Fourier transform (FT) of the source brightness
distribution

\begin{equation}\label{1}
  \gamma(P_{1},P_{2})=\gamma_{1,2}=e^{(-i\psi)}\frac{\int\int_{-\infty}^{+\infty} I(\xi,\eta)\exp\{-\,\frac{2\pi i}{\overline{\lambda}}\,(p\,\xi+q\,\eta)\}d \xi d \eta}{\int\int_{-\infty}^{+\infty} I(\xi,\eta)d \xi d \eta}
\end{equation}
where $I$ is the intensity of the source, $\overline{\lambda}$ - the
mean wave length, $(\xi,\eta)$ are coordinates on the source, and
$(p,q)$ are
\[ p=\frac{x_{1}-x_{2}}{R}, \,\,\,\ q=\frac{y_{1}-y_{2}}{R}\]
Here $R$ is the distance from the source to the points $P_{1}$ and
$P_{2}$ in a projective plane, and $(x_{1}, y_{1})$ and $(x_{2},
y_{2})$ are the coordinates of $P_{1}$ and $P_{2}$, respectively.
Since the function $\gamma_{1,2}$ is complex, it has amplitude and
phase.

We can make measurements of the Fourier magnitude from the
second-order intensity correlations with an intensity interferometer
\begin{equation}
|\,\gamma_{1,2}|\,^{2}= \frac{<I_{1}I_{2}>}{<I_{1}><I_{2}>}
\end{equation}
But in so doing, information about a phase is lost. Third-order
correlations provide advantages over second-order correlations for
image recovery by allowing phase information, as reviewed by Lohmann
and Wirnitzer (1984).

We shall consider 1-D signal reconstruction. The extension to 2-D
images is more complicated, but in principle, it is the same as for
1-D signal. We consider the arbitrary 1-D real signal $I(x)$ whose
third-order spatial correlation is given by (Webster et al. 2002,
2003)
\begin{equation}
        I^{(3)}(x_{1},x_{2})= \int I(x)\cdot I(x+x_{1})\cdot I(x+x_{2})dx
\end{equation}
Lohmann and Wirnitzer (1984) define the bispectrum $B(f_{1},f_{2})$
as the FT of $I^{(3)}(x_{1},x_{2})$
\begin{equation}
B(f_{1},f_{2})=\int\int I^{(3)}(x_{1},x_{2})\,exp\,\{-2\pi
i(f_{1}x_{1}+f_{2}x_{2})\}dx_{1} dx_{2}
\end{equation}
The bispectrum of $I(x)$ may be written as
\begin{equation}
B(f_{1},f_{2})=F(f_{1})F(f_{2})F^{*}(f_{1}+f_{2}),
\end{equation}
where * denotes complex conjugation. We shall express the FT of
$I(x)$ in terms of its magnitude and phase as $F(f) = A(f)\cdot
exp\,(i\,\phi(f))$. Analogically, the bispectrum is defined as
$|B(f_{1},f_{2})| \cdot exp\,(i\,\psi(f_{1},f_{2}))$. The
relationship between the Fourier phase and the bispectral phase of
$I(x)$ may be obtained from Eq. (5) as
\begin{equation}\
\psi(f_{1},f_{2}) = \phi(f_{1}) + \phi(f_{2}) - \phi(f_{1}+f_{2})
\end{equation}
This equation can be exploited to reconstruct the Fourier phase of
$I(x)$ from its bispectral phase. Writing the phases in discrete
notation, allows the Fourier phase to be reconstructed, as noted by
Bartlet et al. (1984)
\begin{equation}
        \phi_{k}=\frac{1}{k-1}\sum_{i=1}^{k-1}\,(\phi_{i}+\phi_{k-i}-\psi_{i,\,k-i})\,,\,\,\,
    k=2,...N,
\end{equation}
where $\phi_{0} = 0$, $\phi_{1}$ is arbitrary, $N$ is the length of
the sequence. Having the value of $\phi_{1}$, any other $\phi_{k}$
can be obtained recursively. A condition for this approaches, that
$\phi_{1}$ is arbitrary, follows from the fact that the bispectral
phase (6) is independent of the linear phase component $\alpha f$.
This gives an arbitrary offset in the reconstructed signal.

The normalized intensity ${\overline{I}} = \Delta I/<I>$, when used
in the third-order spatial correlation in Eq. (2), allow the
bispectral phase to be written as
\begin{equation}\
        \psi(f_{1},f_{2})=\pm \arccos\left[\frac{<\overline{I}_{1}\overline{I}_{2}\overline{I}_{3}>}{2\,(<\overline{I}_{1}\overline{I}_{2}><\overline{I}_{2}\overline{I}_{3}><\overline{I}_{3}\overline{I}_{1}>)^{1/2}}\right]
\end{equation}
Note that there is a sign ambiguity in the bipectral phase in Eq.
(8), which results in time reversal indeterminacy. We can take the
appropriate sign in Eq. (8) using the calculated data. The value of
$\phi_{1}$ linearly effects the phase (7); it may be often estimated
from {\it a priory} data. Thus, by using bipectral technique, we can
reconstruct the Fourier phase. When this is combining with the
Fourier magnitude (2) obtained from the second-order intensity
correlations with an intensity interferometer, we have the Fourier
transform of the source brightness distribution (1). An inverse
Fourier transform recovers the source image.

Some experimental results illustrating the techniques developed
above may be found in Sundaramoorthy et al. (1990), Webster et al.
(2002, 2003).



\end{document}